# Irreversible charging caused by energy dissipation from depinning of droplets on polymer surfaces

*Shuaijia Chen[1], Ronald T. Leon[1], Rahmat Qambari[1], Yan Yan[1], Menghan Chen[1], Peter C. Sherrell[1,2],* Amanda V. Ellis[1],* Joseph D. Berry[1]**

[1]Department of Chemical Engineering, The University of Melbourne, Parkville, 3010, Victoria Australia

[2]School of Science, RMIT University, Melbourne, 3000, Victoria, Australia

E-mail:

peter.sherrell@rmit.edu.au; amanda.ellis@unimelb.edu.au; berryj@unimelb.edu.au

*Abstract*

Interfacial energy dissipation during stick-slip motion of a liquid drop on a non-conductive polymer substrate is shown to lead to an irreversible increase in electrical charge. This previously unobserved phenomenon occurs during surface wetting, in contrast to the previously reported charge separation mechanism that occurs during dewetting. Understanding this electrification mechanism will facilitate the design of energy harvesters and aid the development of risk mitigation strategies for electrostatic buildup in liquid flow across a wide range of industrial applications.

*Introduction*

Electrification of materials upon simple contact, friction, and motion has presented a quandary for philosophers and scientists since, debatably, the ancient Greeks [1]. Traditionally, attempts in understanding the processes governing this so called 'contact-electrification' phenomena have focused on solid-solid material interfaces, due to the ease of fabrication and testing. However, the same electrification phenomena is also observed during motion of solid-liquid interfaces, playing a key role in applications across inkjet printing, energy harvesting, mechanical sensing, risk mitigation, electrospray coating, and microfluid transportation [2-5]. However, despite the importance of electrification at the solid-liquid interface, it remains poorly understood due to challenges in experimentation precision within measurement systems.

In this Letter we have developed a rigorous experimental protocol to probe electrification at solid-liquid interfaces and demonstrate that interfacial energy dissipation from local pinning and depinning of the contact line (air-solid-liquid interface) can play a key role in solid-liquid electrification. We present a protocol to characterise and correlate droplet motion to solid-liquid electrification on a polytetrafluoroethylene (PTFE) substrate, enabling the deconvolution of reversible and irreversible charging events. We show that the irreversible charging events are linked to "stick-slip" motion of the contact line upon depinning from surface asperities, suggesting that energy dissipation associated with depinning may lead to local ionisation and radical formation. These results pave the way for the future design of surfaces with controlled electrification.

Prior key studies have investigated charge transfer between liquids and solids using slide electrification [6-9]. Slide electrification of liquid drops occurs when a drop slides down a tilted plate and the hydrophobic surface is electrically charged [10-13]. The sign and magnitude of this charge can be tuned depending upon surface properties (conductivity, permittivity, surface chemistry, surface energy, and thickness of dielectric layer) [7,8], and liquid properties (viscosity, salt concentration or conductivity, and pH) [14]. The influence of liquid properties has also been investigated in other wetting configurations including the Wilhelmy plate apparatus [15-17].

The prevailing model of solid-liquid electrification, termed "the charge separation model", is predicated on the formation of the electric double layer (EDL) within a liquid electrolyte at the solid-liquid interface. The EDL consists of an adsorption layer of bound surface charges shielded by a diffuse layer of counter charges with a certain Debye length [9,11,18-25]. While the contact line recedes, the EDL separates at the edge of the contact line, the surface charges are deposited on the dewetted surface, and the counter charges remain within the liquid droplet. The deposited charges are adsorbed by the hydrophobic surfaces and lead to the hydrophobic surface being negatively charged [26-31,32].

To date, experimental configurations mainly consider discrete drops moving over surfaces and consequently it is difficult to isolate the effect of wetting and dewetting contact lines. Here, we use a sessile-drop goniometry configuration [33], in combination with charge measurements, to examine effects of wetting and dewetting contact line motion on PTFE. We show that irreversible charging occurs due to "stick-slip" contact line motion during surface wetting, proposed to be driven by interfacial energy dissipation. This discovery opens up exciting new possibilities to either promote or mitigate charging (depending upon application) *via* control of liquid contact line motion.

*Methods*

Polymer samples used consisted of commercial polytetrafluoroethylene (PTFE), thickness 500 μm (Swift Supplies, Australia), cut to size 4×4 cm. One side of each PTFE sheet was sputter coated (Quorum Q150T ES, Quorumtech, UK) with a 3 nm chromium (Cr) layer followed by a 30 nm gold (Au) layer. The electrode side (not in contact with liquid) of each PTFE sample was covered with Kapton (polyimide) tape (50 μm thick, RS Components, Australia) for insulation. This side of the PTFE was then adhered to a glass substrate (Premiere, Microscope Slides, 2” × 3” 1.00 mm) with double-sided tape (3M) to ensure a flat contact surface. Before liquid contact, two cleaning protocols were considered. For the first, the PTFE sample was placed onto the Kapton tape insulated stage (Figure S1) and cleaned with Kimwipe(TM) cleaning tissue (Kimberley-Clark) to promote droplet pinning. In the second, nitrogen gas was blown onto the surface for ~2 min before being placed onto the stage.

To begin the experiment, a 20 μL droplet of deionised water (Milli-Q, resistivity 18.2 MΩ·cm) was dispensed onto the PTFE surface, and the needle (ø = 0.72 mm) position was manually adjusted to the droplet centre. A programmable syringe pump (NE-1000, NewEra Pump System, USA) was used to control the droplet volume by adjusting the volumetric flow-rate $Q$ (Figure S1). Each trial consisted of 10 cyclic dispensing (wetting) and withdrawing (dewetting) stages for each droplet, with a change in volume ($\Delta V$) of 150 μL at a constant $Q$ (300, 600, 900 or 1200 μL/min). A 60 s pause between each stage was used to allow for droplet and charge equilibration [24]. Droplet contact line motion was captured *via* a CMOS camera (Basler acA1920-150um, BASLER, Germany) at 25 fps or 100 fps, and electrical charge was measured *via* an electrometer (KEITHLEY-6514, Tektronix, USA) connected to a data acquisition system (DAQ, NI-9223, National Instruments, USA). The electrical measurements, performed in the 0 - 20 nC range, provide a nominal resolution of 10 fC. The typical measured charge, of order 0.1 - 1 nC, is sufficiently above this measurement resolution to ensure reproducible and consistent measurements with low signal-to-noise ratio. A small drift in charge occurred over long timescales due to external environmental factors, with no appreciable effect on the change in charge measured during each wetting and dewetting cycle (Figure S2). Contact angle and contact radius data were extracted from recorded videos using custom Python code.

*Measuring & Correlating Wetting and Surface Charge*

To systematically study the role of droplet motion on solid-liquid charging, 20 μL droplets were deposited on the surface of a PTFE film after wiping with Kimwipe(TM) cleaning tissue (Figure 1a). These droplets then underwent cyclic testing via sequential wetting and dewetting stages (Figure 1b), with the contact line positions and angles measured *via* camera, and simultaneous measurements of charge ($q$) with a Keithley 6514 electrometer connected to the Au/Cr electrode. The typical change of contact area of the droplet at $Q$ = 1200 μL/min shows the expected increase in contact area through each wetting stage, and decreases in contact area in each dewetting stage (Figure 1c). No adaptive wetting was observed on the PTFE samples [34, 35], with a typical measurement showing an advancing contact angle of 100°–105° and a receding contact angle of ~80° (Figure S3). The simultaneous charge measurement of the same droplet is presented in Figure 1d. Notably, there is significantly higher $\Delta q$ observed in the first wetting stage, where charge increases from 0 nC to 4.1 nC. The first dewetting stage, and all subsequent cycles from 2 to 10, only show a $\Delta q$ of 0.9 nC (between ~3.2–4.1 nC). Further, when the droplet is not moving (during the 60 s pause) the charge remains constant.

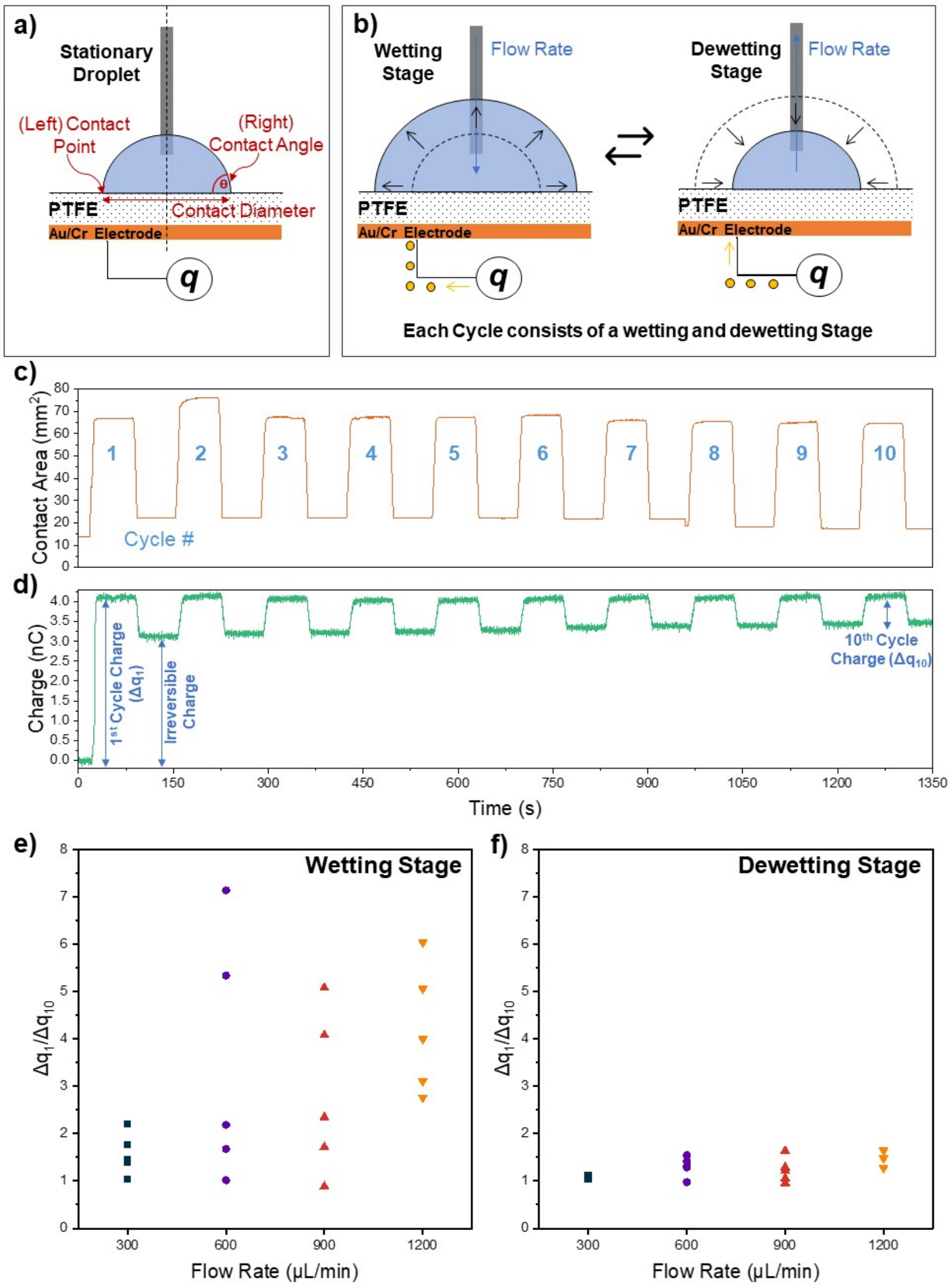

**Figure 1. | a.** Schematic of droplet measurement, labelled with key measured parameters, contact point position, contact angle, and contact diameter, used for determining the contact area; **b.** an experimental cycle, consisting of a wetting stage whereby the droplet is grown by injection of liquid via syringe and a dewetting stage, whereby the droplet is shrunk by extraction of liquid via syringe; **c.** typical contact area measurement calculated from the contact diameter (in a.) during a 10 cycle experiment using a 150 µL drop on PTFE at flow-rate, $Q$ = 1200 µL/min, each increase (decrease) in contact area corresponds to a wetting (dewetting) stage; **d.** simultaneous charge measurement during the 10 cycles; **e.** the ratio of the change in charge for the 1$^{st}$ cycle relative to the change in charge for the 10$^{th}$ cycle as a function of different $Q$ for e) wetting and f) dewetting.

Using these charge measurements as a baseline, irreversible charging is defined as a cycle where $\Delta q_{\text{wetting}} \neq \Delta q_{\text{dewetting}}$, with reversible charging defined as occurring in a cycle when $\Delta q_{\text{wetting}} \approx \Delta q_{\text{dewetting}}$. Irreversible charging was predominantly observed in the first wetting stage (Figure 1d), whereby $\Delta q$ is significantly higher for the first wetting cycle (from 0 nC to 4.1 nC) compared to the tenth wetting cycle (from 3.2 nC to 4.1 nC) (Figure 1d), despite the wetted area change remaining relatively consistent across all cycles (Figure 1c, Figure S4). At cycle numbers ≥2 reversible charging was consistently observed to arise independent of $Q$ from 300–1200 µL/min. Thus, taking the ratio of $\Delta q_{1,wetting}$ and $\Delta q_{10,wetting}$ for each individual experiment (see definitions in Figure 1d), the influence of sample-to-sample variance can be removed and the influence of *Q* on irreversible charging can be elucidated (Figure 1e). Here, increasing *Q* leads to a greater chance for irreversible charging to occur. Interestingly, this phenomenon is only observed in the first wetting cycle, with the equivalent ratio for wetting ($\Delta q_{1,dewetting}/\Delta q_{10,dewetting}$) showing a near convergence to 1 (Figure 1f). At *Q* of 300, 600, 900, and 1200 µL/min (Figure S5) the average reversible charge during wetting was 0.8 ± 0.1, 0.6 ± 0.1, 0.7 ± 0.2, and 0.7 ± 0.1 nC respectively, with the average reversible charge during dewetting being 0.8 ± 0.1, 0.6 ± 0.1, 0.6 ± 0.2, and 0.7 ± 0.1 nC respectively. The overall average reversible charge for distilled water droplet motion on PTFE was 0.69 ± 0.13 nC across all samples.

The independence of $\Delta q$ with *Q* during reversible charging (cycles ≥ 2) at later wetting cycles suggests that the charging in the reversible regime is due to the formation of an electric double layer at the solid-liquid interface and subsequent ion adsorption, with no influence due to contact line velocity (Figure S6).

Quantifying the charge adsorbed during reversible charging is achieved by considering the observed change in wetted area, $\Delta A$ = ~50 mm$^2$ (Figure 1c) and $\Delta q$ = ~0.9 nC (Figure 1d) results in an adsorbed surface charge density, σ = ~18 µC/m$^2$. This is consistent with Li et al. [7,14] and Stetten et al.'s [9] observation of adsorbed charge density on perfluorodecyltrichlorosilane (PFOTS) surfaces of σ = 10-15 µC/m$^2$, and Hinduja et al. 's [32] estimates of σ = 20 µC/m$^2$ and σ = 45 µC/m$^2$ on octadecyltrichlorosilane (OTS) and PFOTS coated glass surfaces, respectively.

*Pinning and Depinning*

To investigate the irreversible charging observed predominantly in the first wetting cycles, 380 first wetting stage experiments were performed with water droplets on PTFE surfaces. The occurrence and magnitude of irreversible charging were not consistent for all experiments, with a wide range of change in charge $\Delta q$ during the first cycle. The irreversible charging was observed to correlate with discrete pinning-depinning events. These events are schematically shown in Figure 2a, generating a droplet interface motion termed "stick-slip" motion, and typically arise from regions of different surface energy or roughness (defects, contaminants, topography) [36]. Here, it was observed that wiping the PTFE surface with Kimwipe$^{(TM)}$ cleaning tissue led to a significant increase in observed pinning and depinning events.

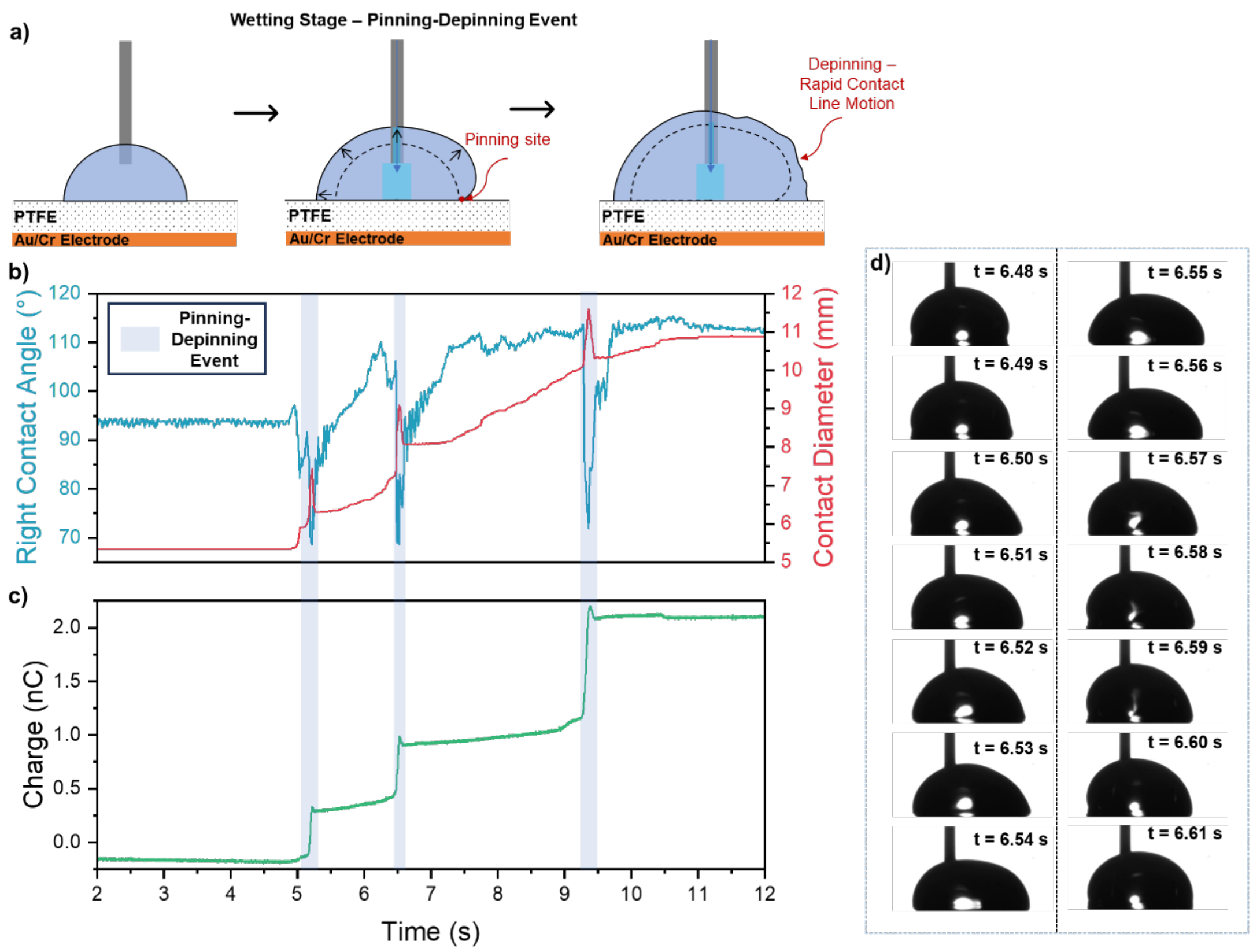

**Figure 2. Pinning-Depinning Events| a.** Schematic of droplet motion during a wetting stage when a pinning events occurs, (1) a standard, symmetrical droplet; (2) the right contact point pins; (3) the continued injection of water overcomes pinning, leading to rapid contact line motion and a decreased contact angle; **b.** Contact angle measurement (blue) with corresponding variation of the liquid droplet's contact diameter (red) versus time during the first wetting stage; **c.** Corresponding charge measurement (green) during the wetting stage shown in b (The blue shaded regions refer to "stick-slip" motion arising from pinning-depinning events); and **d.** Captured droplet images of the liquid drop's motion during the period of the second pinning-depinning event (second shaded region highlighted in b,c).

Figure 2b, shows the variation of contact angle (blue) and contact diameter (red) during 3 pinning/depinning events on an example 1$^{st}$ wetting stage at $Q$ = 1200 µL/min. The contact line exhibits stick-slip behaviour (shaded blue regions), where there are discrete, sharp jumps in contact point position accompanied by a sharp decrease in contact angle (>20°). These contact line pinning-depinning events directly correlate to sharp jumps in the measured charge accumulation (Figure 2c, green). Of note is that in this measurement the left contact point does not move significantly throughout the cycle, whereas the right contact point pins and depins three times (Figure 2d, Figure S8, Movie A). Within these experiments, stochastic pinning-depinning events of the droplet were observed in ~55% of the cycles, with multiple events often occurring within one cycle. These pinning-depinning events were observed to be universally correlated to the occurrence of irreversible charging (Figure 2b,c).

The data from 300 first wetting stages is visually represented as a contour map (Figure 3 a,c) where rapid changes from yellow to purple denote depinning events and gradual changes from yellow to orange reversible charging. The highest observed total charge (~6 nC) occurs when multiple contact line jump events occur within the single run. Replotting the graph with a colour scale between 3 and 6 (Figure S9) shows the heterogeneity of the data within this irreversible charging regime.

In total, over 670 pinning-depinning events were observed (see Figure S10 for details). Figure 3d shows the distribution of the corresponding charge accumulation from each pinning-depinning event. The data is well fitted by a log-normal distribution with median $\Delta q_j$ = 0.31 nC. The maximum irreversible charge we observed from a single pinning-depinning event was 2.3 nC, ~3 times the average charge generated via reversible charging (0.69 ± 0.13 nC).

The log-normal distribution of charging events is consistent with the distribution of asperity size associated with a surface of random spacing between asperities [37]. This suggests that the magnitude of $\Delta q_j$ is correlated with the size of the pinning-depinning event. To probe the link between pinning-depinning events and irreversible charging, different cleaning methods were employed to produce pristine PTFE surface prior to measurement. The most effective approach was found to be simple nitrogen gas cleaning for 120 s. Surfaces measured using this method demonstrated minimal pinning-depinning events (~16% probability over 80 experiments, Figure 3b, Figure S11).

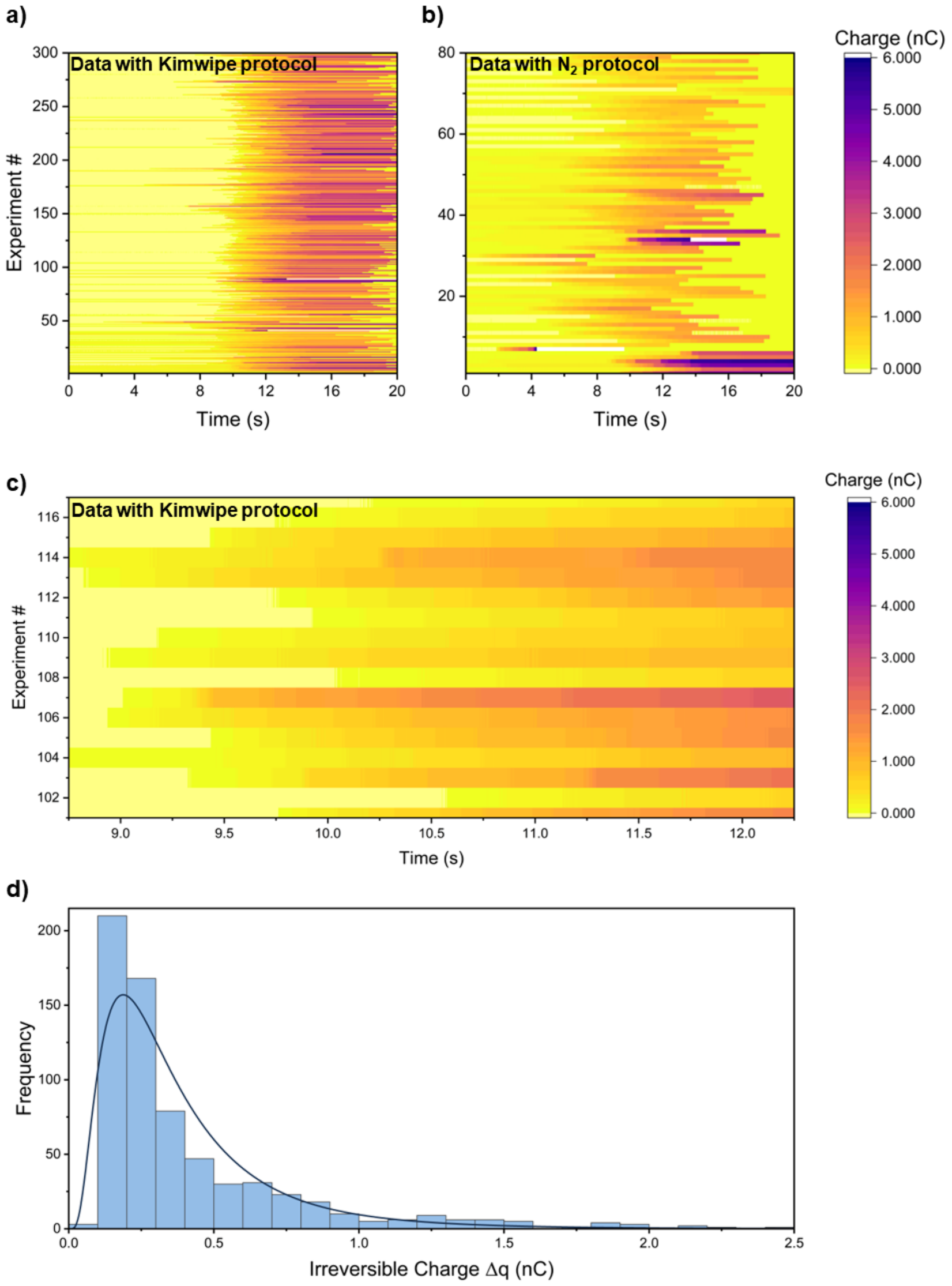

**Figure 3. a.** Contour map of charge showing individual experiments with pinning-depinning events on Kimwipe™ cleaned PTFE samples compared to; **b.** $N_2$ cleaned PTFE sample which has significantly fewer pinning-depinning events; **c.** Expanded region of a) showing variability across experiments; and **d.** measured charge distribution corresponding to 670 individual pinning-depinning events observed on Kimwipe-cleaned PTFE samples, with log-normal fit (blue line, $\mu = -1.175$, $\sigma = 0.706$)

## *Interfacial Energy Dissipation*

The pinning and depinning of a droplet contact line moving on a surface is known to correspond to significant energy dissipation *via* capillary waves, viscous flow and local heat generation [38-41]. The energy dissipated *via* each pinning-depinning event can be estimated as $E_i \sim \sigma A_{CV}$ [42], where $\sigma$ is the air-liquid interfacial tension (72 mN/m for air-water) and $A_{CV}$ is the change in wetted area that occurs during the depinning of the interface. This can be estimated as $A_{CV} \sim l_j^2$, where $l_j$ is the length scale of the contact line jump during the depinning event [42].

For the second depinning event in Figure 2b, $l_j$ =1.86 mm, giving an energy dissipation estimate of $E_i = 2.5\times10^{-7}$ J. This can be compared to the change in electrical potential energy $E_q = 0.5\Delta q_j V$ caused by the depinning event, where $\Delta q_j$ is the measured change in charge for the depinning event and $V$ is the electrical potential which can be calculated *via* the capacitance $C$ with $V = \Delta q_j/C$. The capacitance is given by $C = (A\varepsilon_r\varepsilon_0)/d$, with a drop contact area, $A$, of 84 mm$^2$, material thickness, $d$, of 500 µm, and relative permittivity, $\varepsilon_r$, of 2.05. The $\Delta q_j$ for the second depinning event in Figure 2c is ~ 0.52 nC, giving an estimate of the change in electrical potential energy $E_q = 0.5\Delta q_j^2/C = 4.4\times10^{-8}$ J for this event. For these depinning events (and all other observed events), $E_q < E_i$, suggesting that a portion of the energy dissipated from a depinning event gives rise to charge accumulation on the PTFE material, analogous to the charge generated *via* the motion of two contacting solid surfaces, which dissipate energy due to interfacial friction leading to charge transfer [4].

In the work presented here, up to ~18% of the energy dissipated leads to irreversible charge generation, significantly higher than mechanical-to-electrical conversion efficiency for most solid-state frictional energy harvesters [43]. The remaining dissipated energy is lost through localised heating, viscous dissipation, and capillary waves [38-41] which can be observed in the oscillations of the contact angle from $t = 6.6$ s to $t = 7$ s in Figure 2b (blue, see also Figure S12). These oscillations are also schematically shown in the description of pinning-depinning events in Figure 2a.

*Discussion of Irreversible Charging*

Within literature charge transfer from droplet motion is typically studied using motion to create either fast linear flow (*via* addition of a slope or gradient) or droplet impact [7]. These approaches don't allow study of individual charging events and thus the role of energy dissipation has been overlooked. To develop an understanding of the observed phenomena, borrowing from solid-solid interfacial electrification may bear merit, particularly for polymer systems, where charging events have been tied to covalent bond scission (i.e. energy dissipation) at the interface [44, 45]. In this model, bond scission leads to radicals (from homolytic bond scission) and ion-terminated oligomers (from heterolytic bond cleavage), with charge transfer observed based off the relative affinities of each contact surface for the generated charges. In a solid-liquid model, the dissipated energy and localised heat generation may lead to radical formation and ionisation of both surface species and water molecules, measured as irreversible charge stabilised within the droplet [46]. Indeed, Chen et al., [46] have shown OH radical production by flowing water through a glass/polydimethylsiloxane (PDMS) microfluidic channel. On subsequent wetting-dewetting cycles, these ions and radicals internalised within the water droplet are then measured as reversible charge *via* induction over the measurement electrode. Alternative models describe electron transfer occurring at the solid-liquid interface [47], which are unlikely to occur in this case due to the lack of compressive force leading to the electronic orbital overlap required.

For irreversible charging to occur, species must go from a neutral to a charged state and be stabilised. This suggests that charged species bound to the surface of the polymer must be generated, as moving charges within the droplet would lead to larger reversible charging through ion motion. These charged species, are likely to be a combination of mechano-radicals and mechano-ions [45, 48], shown to stabilise each other in solid-solid frictional charging systems [49], however further study is needed to prove this effect.

*Conclusion*

Here, a new cause for charge generation in solid-liquid interfaces is shown, hitherto undiscovered in the literature, through interfacial energy dissipation *via* local contact line jumps of a droplet on a PTFE surface. This charging mechanism is irreversible, and occurs at the contact line during wetting, in contrast to the charge separation from the electric double layer observed to occur at the contact line during dewetting [24, 32]. The discovery of these charging events, combined with the extensive existing knowledge on engineering pinning-depinning events on surfaces, presents exciting opportunities for controlling and manipulating these charging phenomena. Depending on specific application, these findings can be utilized to either exploit or mitigate charging events, providing a pathway to a new generation of surface coatings, with significant implications for the design of energy harvesting devices and risk mitigation strategies for the chemical industry involving handling of flammable liquids.

*Acknowledgements*

PCS acknowledges support from RMIT University *via* the RMIT Vice-Chancellors' Fellowship Scheme (2023). JDB is the recipient of an Australian Research Council Future Fellowship (FT220100319) funded by the Australian Government. This work was performed in part at the Materials Characterisation and Fabrication Platform (MCFP) at the University of Melbourne.

**Supporting Information**

**Irreversible charging caused by energy dissipation from depinning of droplets on polymer surfaces**

*Shuaijia Chen, Ronald T. Leon, Rahmat Qambari, Yan Yan, Menghan Chen, Peter C. Sherrell,* Amanda V. Ellis,* Joseph D. Berry**

S. Chen, R. T. Leon, R. Qambari, Y. Yan, P. C. Sherrell, A. V. Ellis, J. D. Berry,
Department of Chemical Engineering, The University of Melbourne, Parkville, 3010, Victoria Australia
E-mail: amanda.ellis@unimelb.edu.au ; berryj@unimelb.edu.au

P. C. Sherrell,
School of Science, RMIT University, Melbourne, 3000, Victoria, Australia
Email: peter.sherrell@rmit.edu.au

## 1. Measurement variability

The movement of liquid droplets on the PTFE surface was arbitrary and difficult to control. For an ideal wetting situation, the needle remains in the centre of droplet and the entire contact line moves. In contrast, for some wetting scenarios, either the left or right contact point is pinned to the needle without any significant movement and the distance of the unpinned side movement is approximately doubled compared to an ideal situation. Sometimes, the droplet wets the surface and elongates in another direction that is parallel to the camera's view, therefore it was difficult to interpret the motion of the droplet from the recorded videos. In approximately 80 first wetting trials (out of 380 total trials) for the Kimwipe™-cleaned PTFE surface, the droplet detached from the needle and split into multiple smaller droplets.

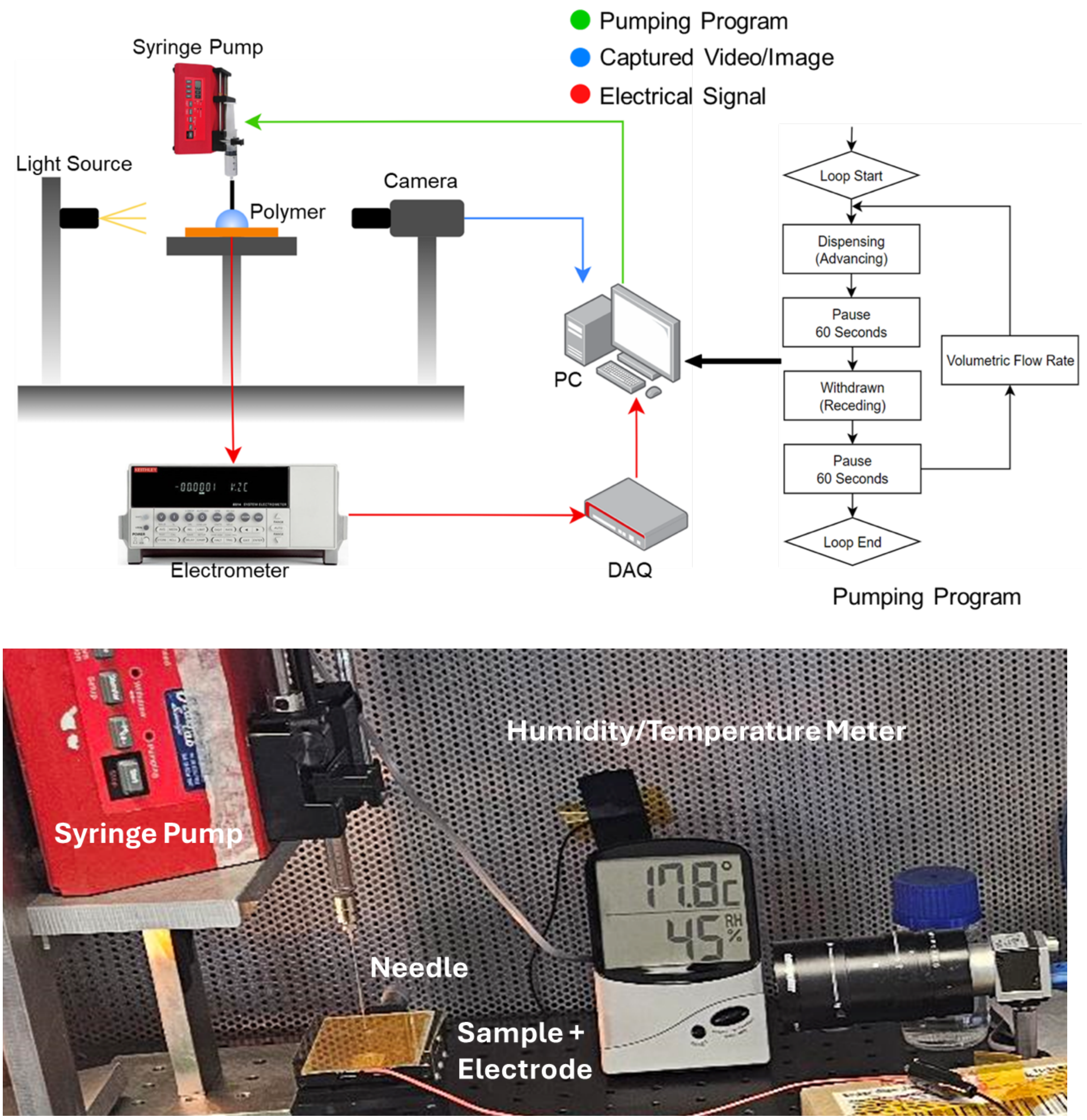

**Figure S1. a.** Schematic diagram of the experimental set-up for simultaneously measuring contact angle changes with surface charge using an adjustable liquid pumping program. **b**. Close up view of the sample stage and experimental set-up with the PTFE sample being placed onto the Kapton tape insulated stage and connected to the measurement equipment.

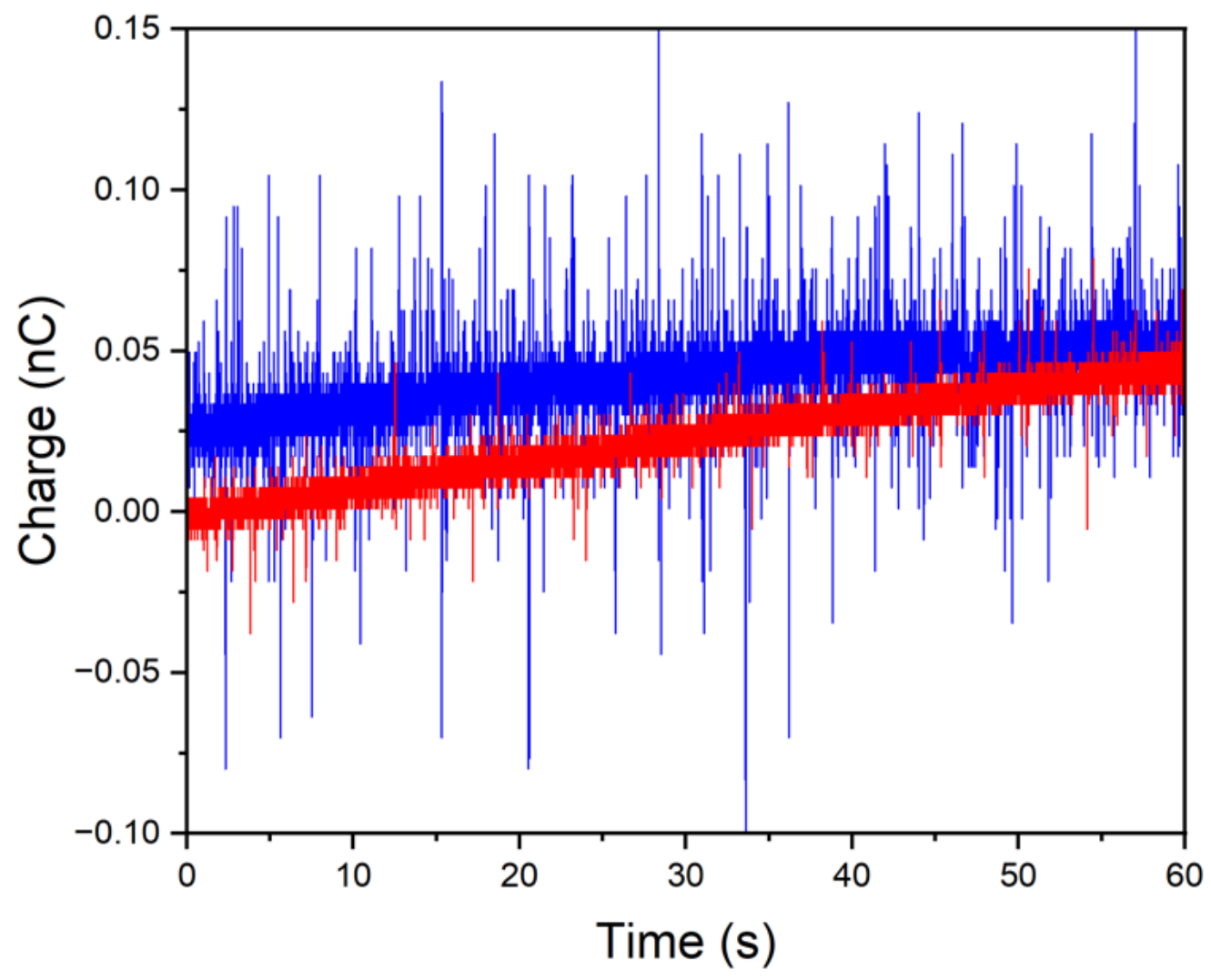

**Figure S2**. Characterisation of background charge drift in experiments. The measurement in blue indicates an experiment where the drop remained static on the drop surface, and the measurement in red is an experiment without a drop on the surface.

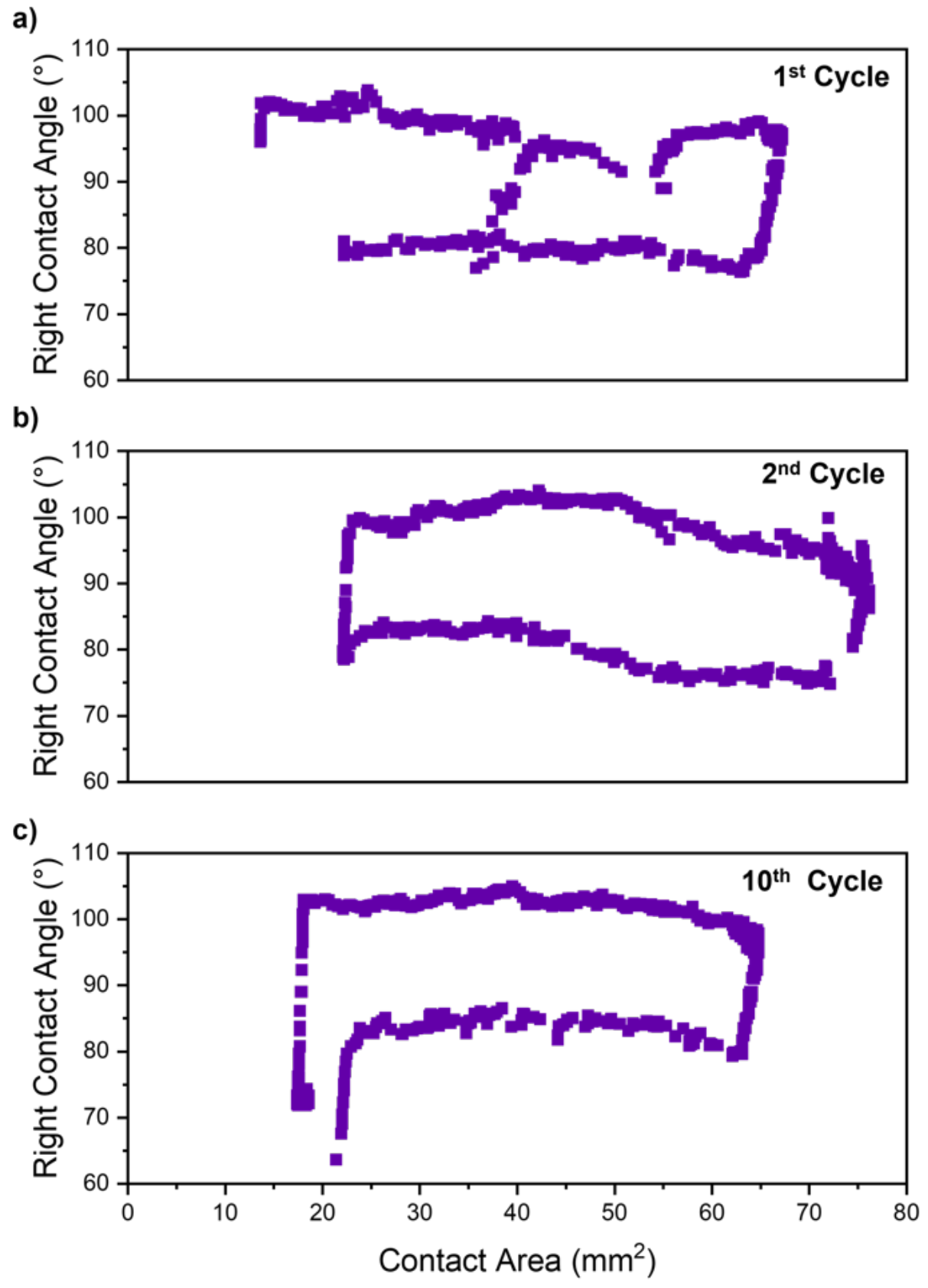

**Figure S3**. Comparison of the contact angle as a function of drop contact area ($mm^2$) for the a) first wetting/dewetting cycle, b) the second wetting/dewetting cycle and c) the tenth wetting/dewetting cycle.

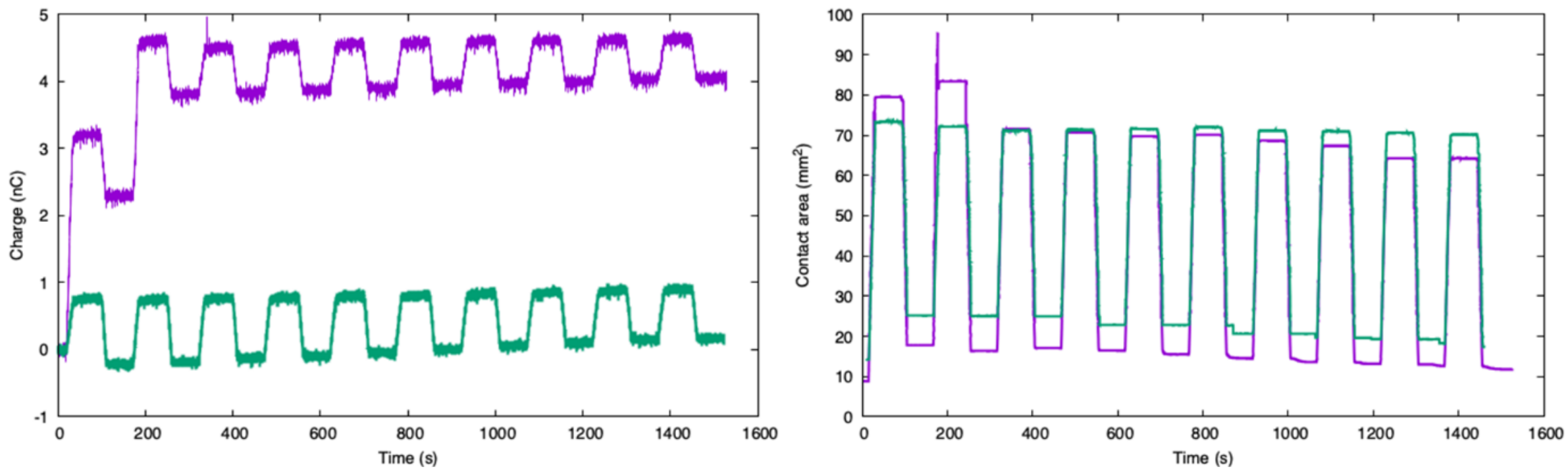

**Figure S4.** Two charge measurements (left) and corresponding change in wetted contact area (right) at flow rate Q = 600 uL/min. The purple line indicates a measurement where depinning events were observed, and the green line indicates a measurement where no depinning events are observed.

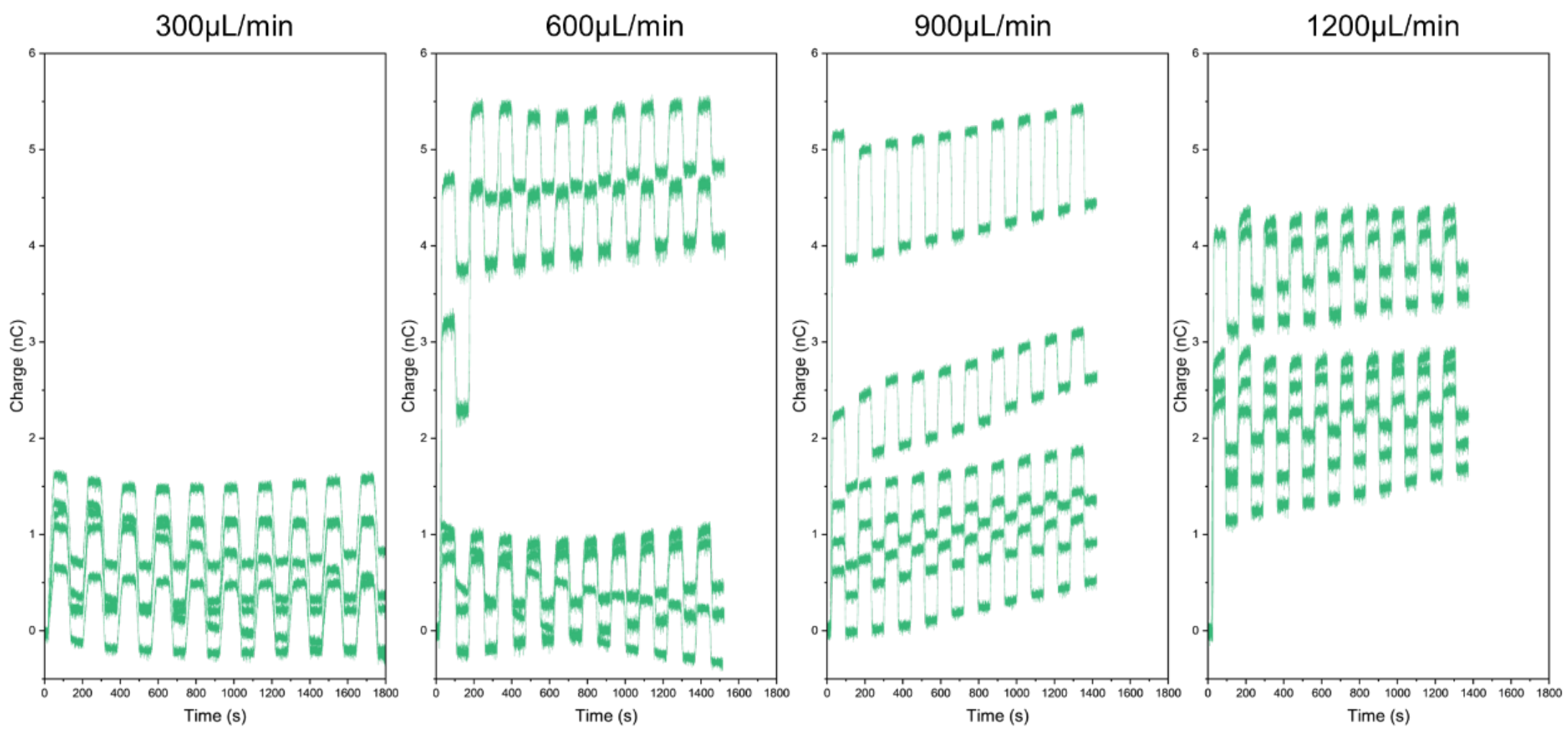

**Figure S5.** Typical experimental results of measured electrical charge signal over 10 wetting-dewetting cycles at various flow rates at 300, 600, 900, and 1200 μL/min.

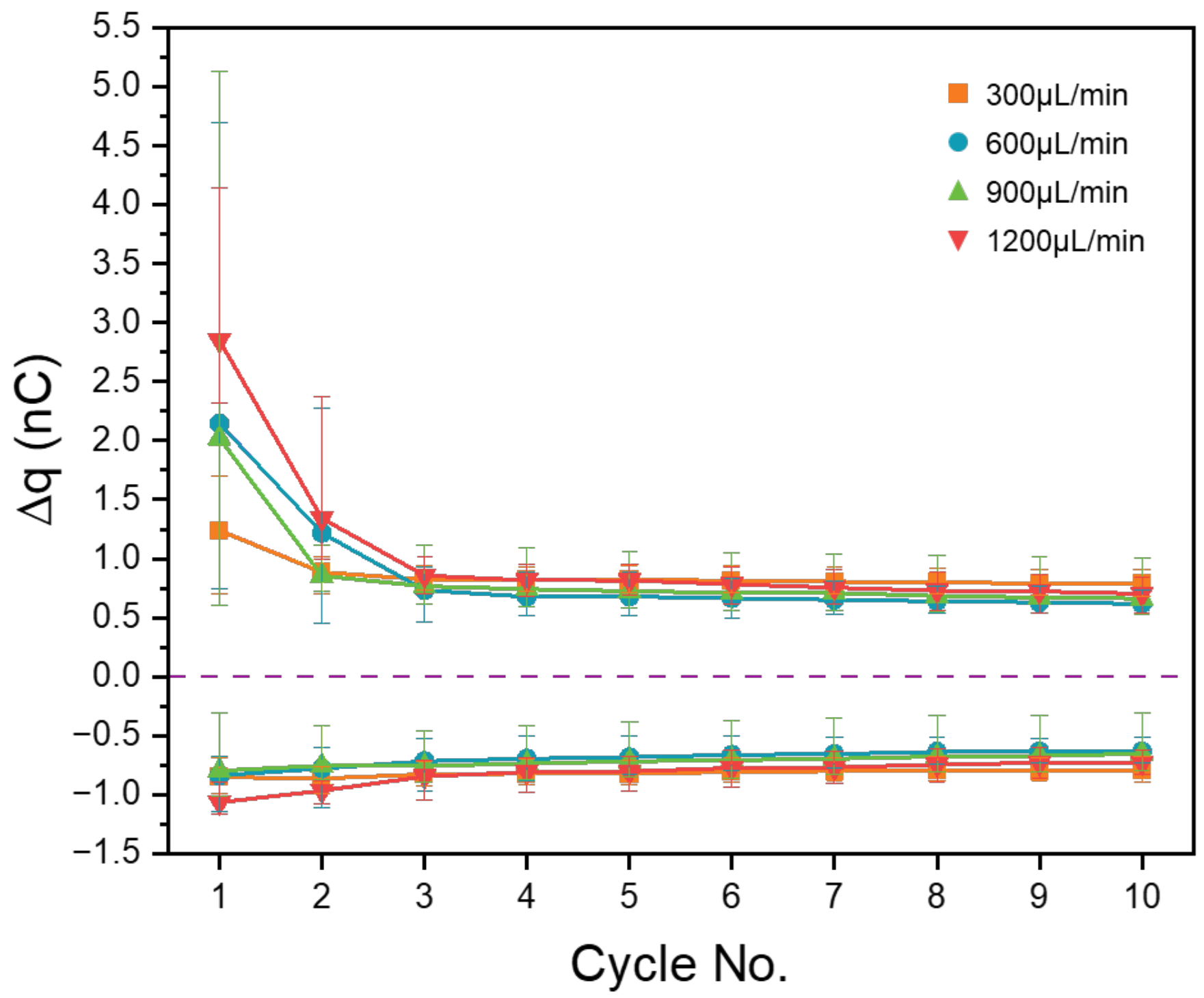

**Figure S6.** The overall correlation between averaged change ($\Delta q$) in charge and various volumetric flow rates ($Q$) with positive values representing the wetting stage and negative values representing the dewetting stage.

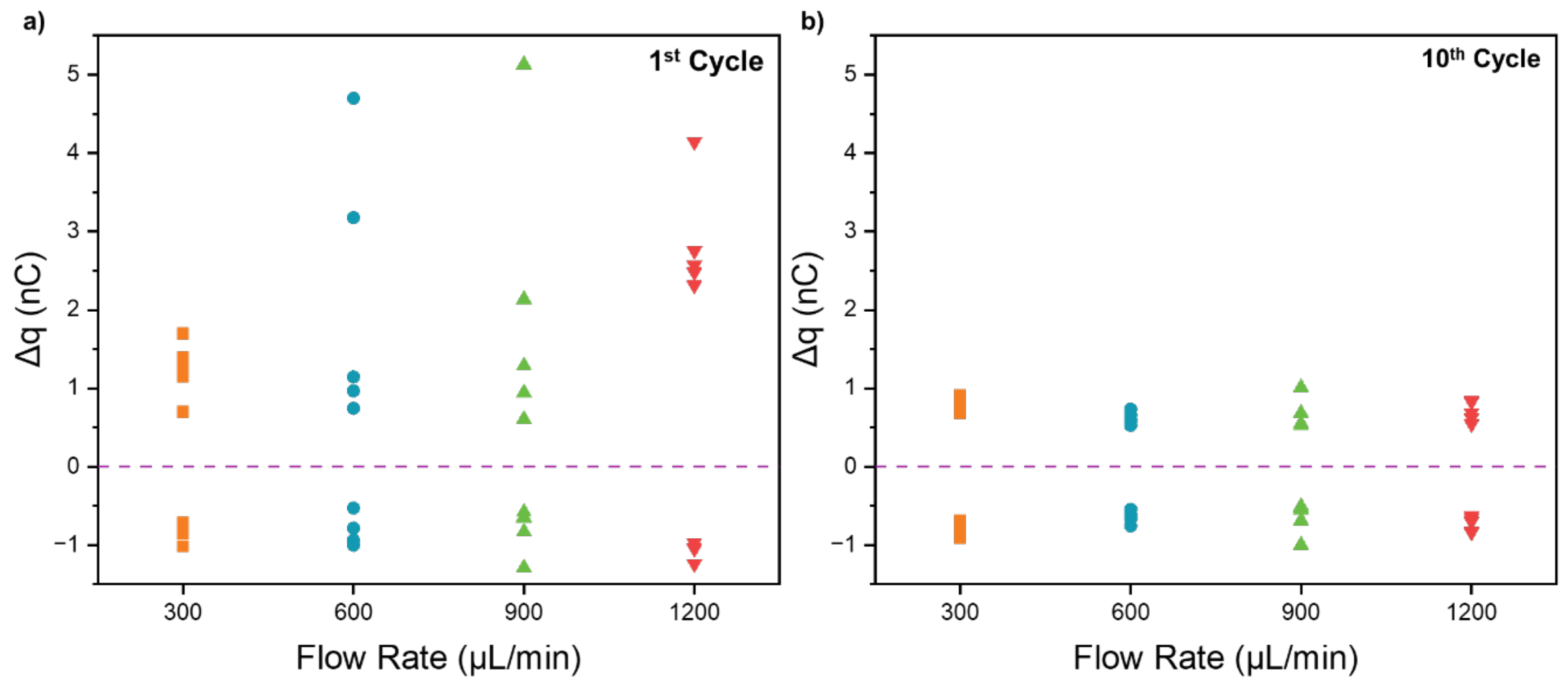

**Figure S7. a,b.** The comparison of the 1$^{st}$ and 10$^{th}$ wetting cycle of change in charge ($\Delta q$) at various flow rates ($Q$) with positive values representing the wetting stage and negative values representing the dewetting stage.

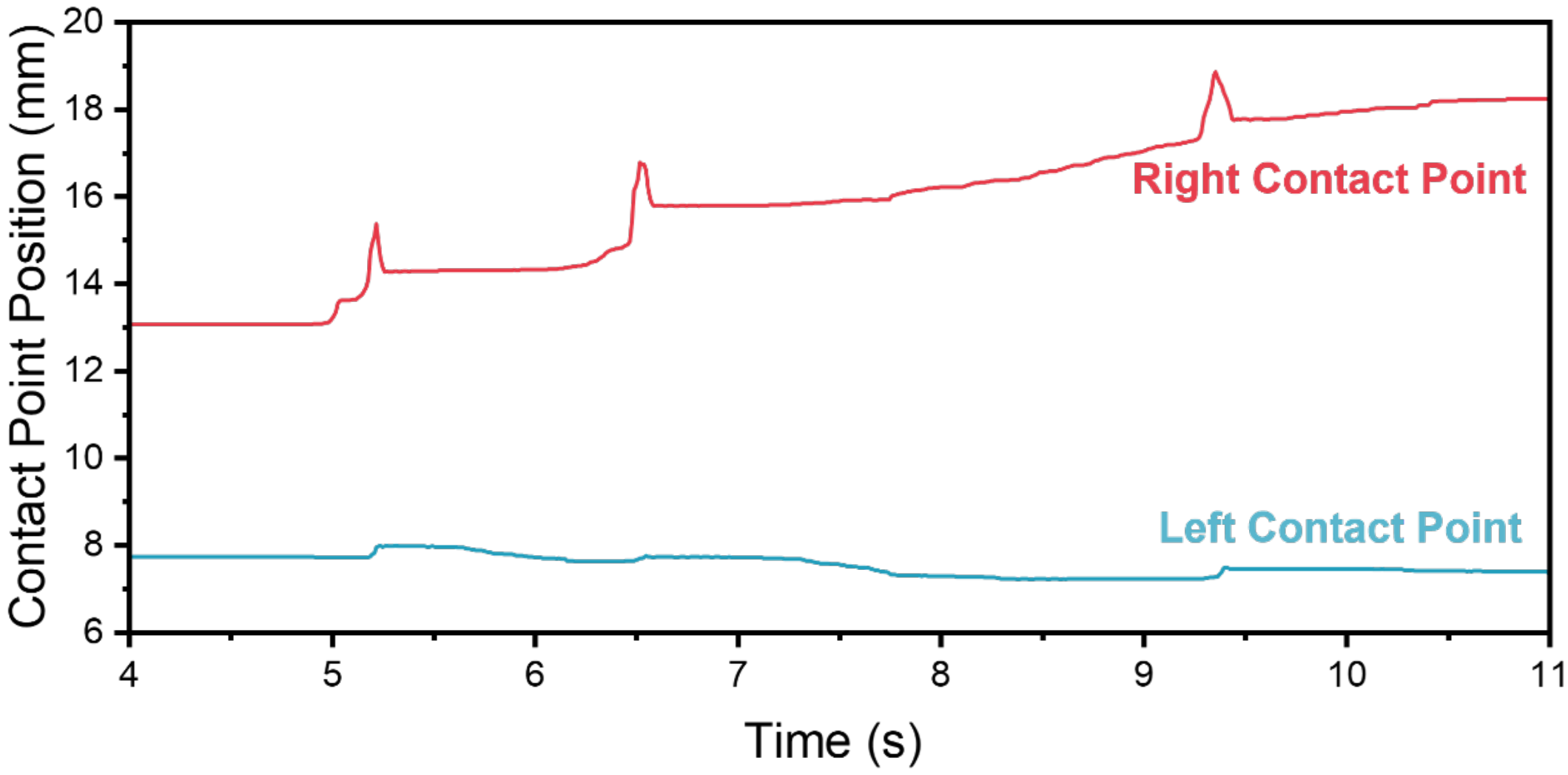

**Figure S8.** The variation of the left and right contact points' position during the first wetting cycle, shown in Figure 2.

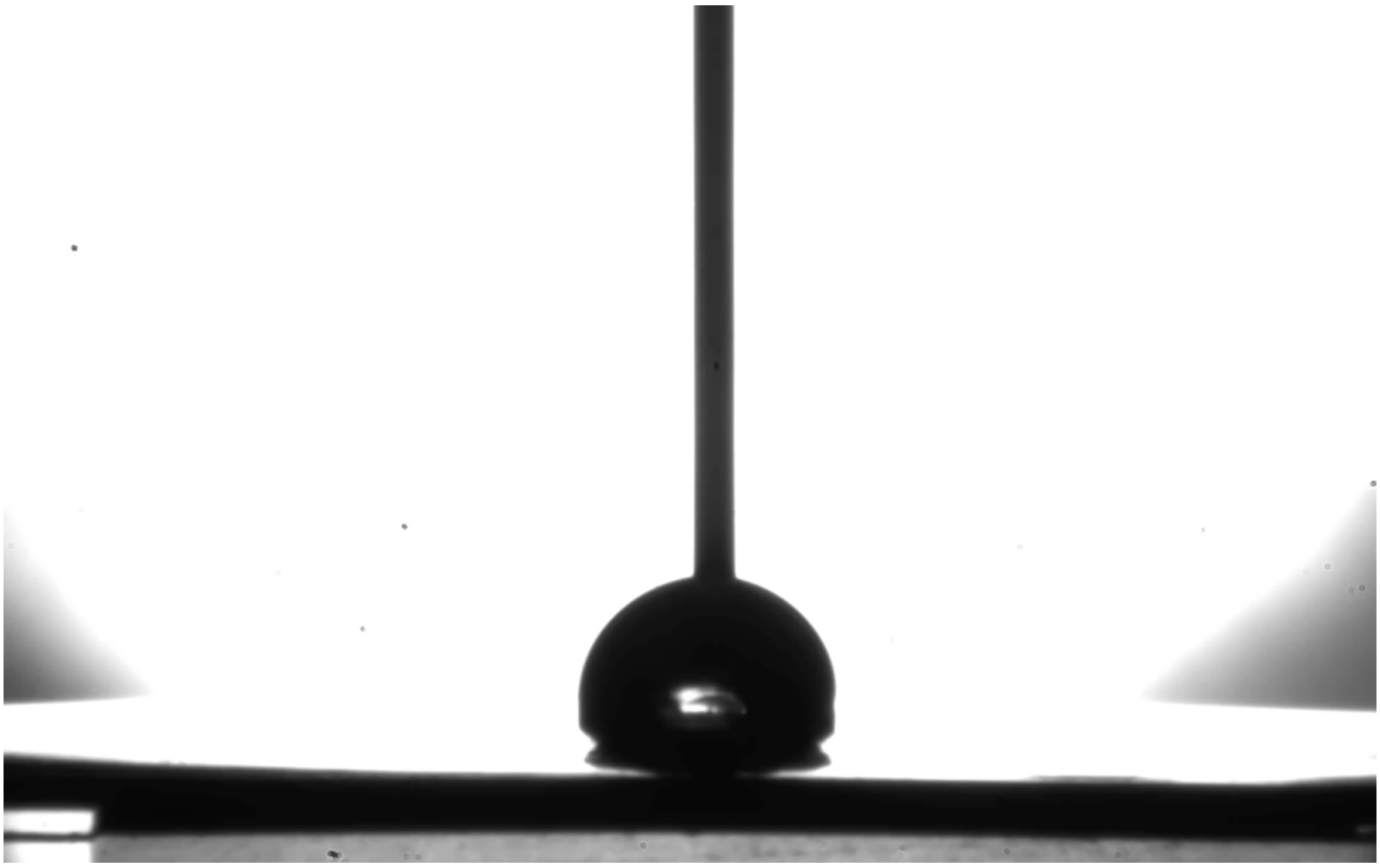

**Movie A.** Recording of droplet motion during a first wetting cycle on Kimwipe™-cleaned PTFE for a flow rate of 1200 µL/min.

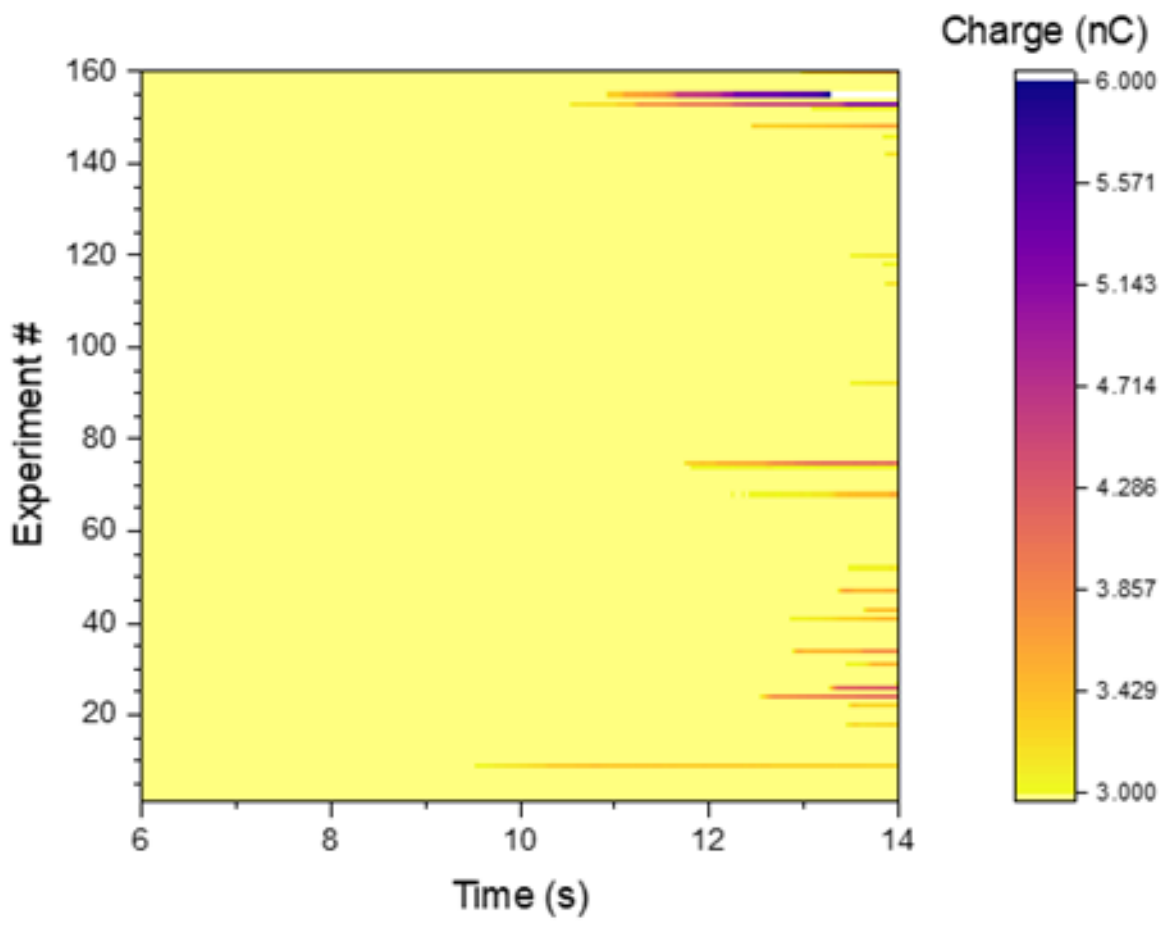

**Figure S9.** Rescaled colour image of Figure 3a showing heterogeneity of jump events, with some runs exhibiting charge over 4 nC.

## 2. Quantification of charge jump events

The charge signal was down-sampled from 500 Hz to 25 Hz to remove noise. The current was then estimated using a sliding window algorithm to determine the local slope of the charge measurement. A 4-point window and a second-order windowed model were used [1]. Following this, the 'findpeaks' function in Matlab was used with a minimum peak separation of 80 to locate local maxima in the current data. A jump event was defined as a peak current greater than 1.2 nC/s. The size of each jump event was defined to be the point either side of the peak value where the current dropped below 1 nC/s. The difference in charge between these two points (blue and pink points in Figure S3 below) was then used to calculate the change in charge for each jump event. Based on the parameters chosen, the minimum measurable charge jump event is ~0.1 nC.

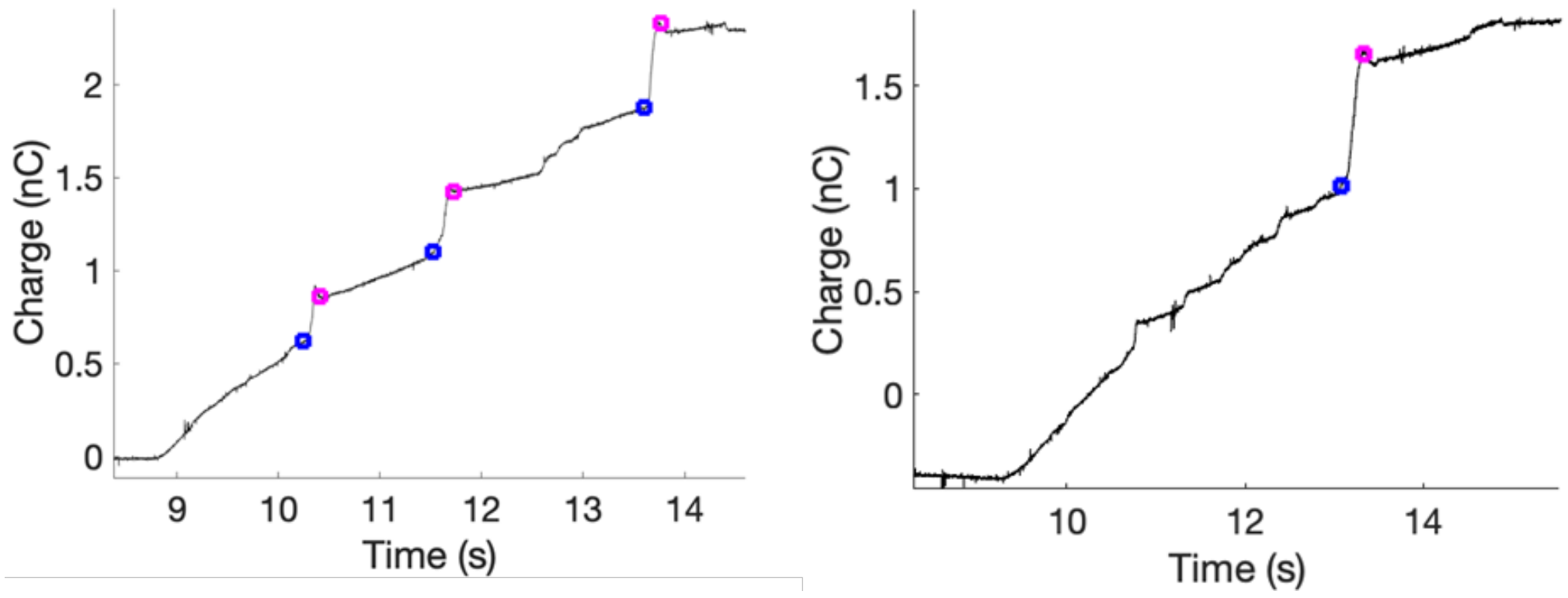

**Figure S10.** Two example first wetting cycles on Kimwipe™-cleaned PTFE, showing the identification of different jump events *via* the Matlab code. The blue symbol indicates the start of a jump event and the magenta symbol indicates the end of a jump event.

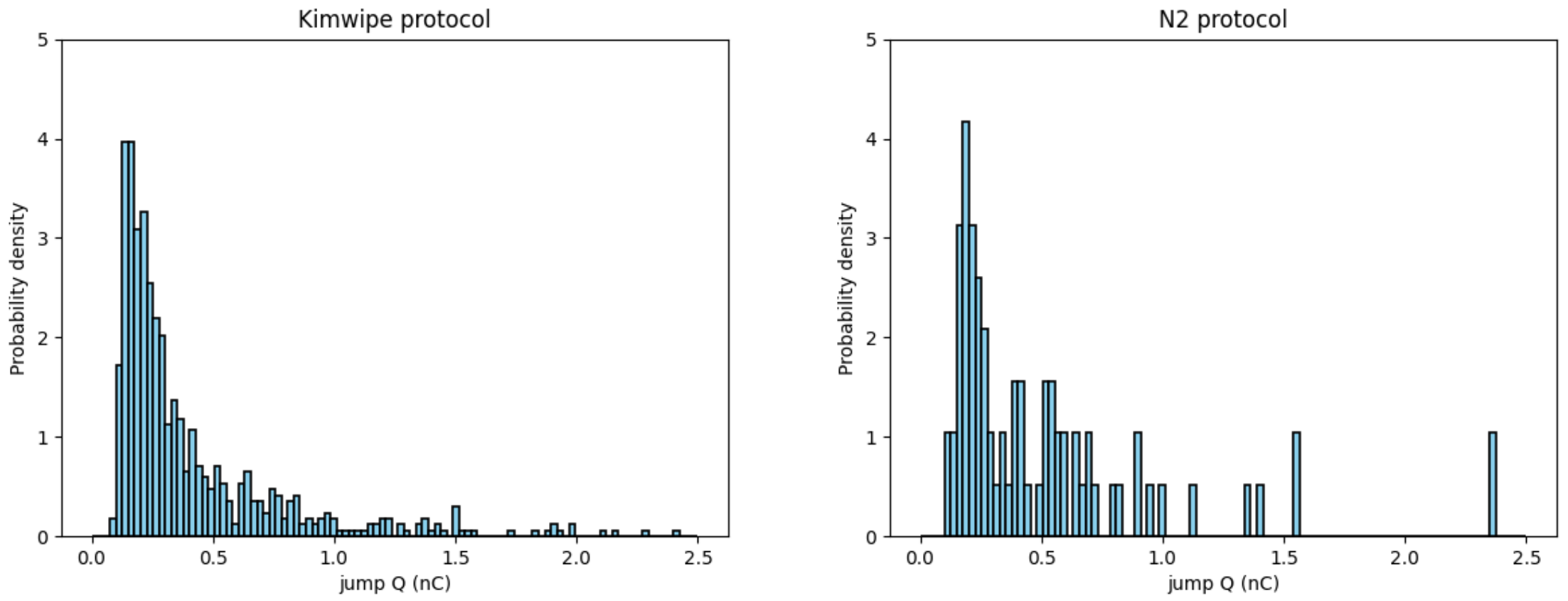

**Figure S11.** Probability density functions of irreversible charge jumps observed for **a.** Kimwipe protol, and **b.** the N2 protocol**.**

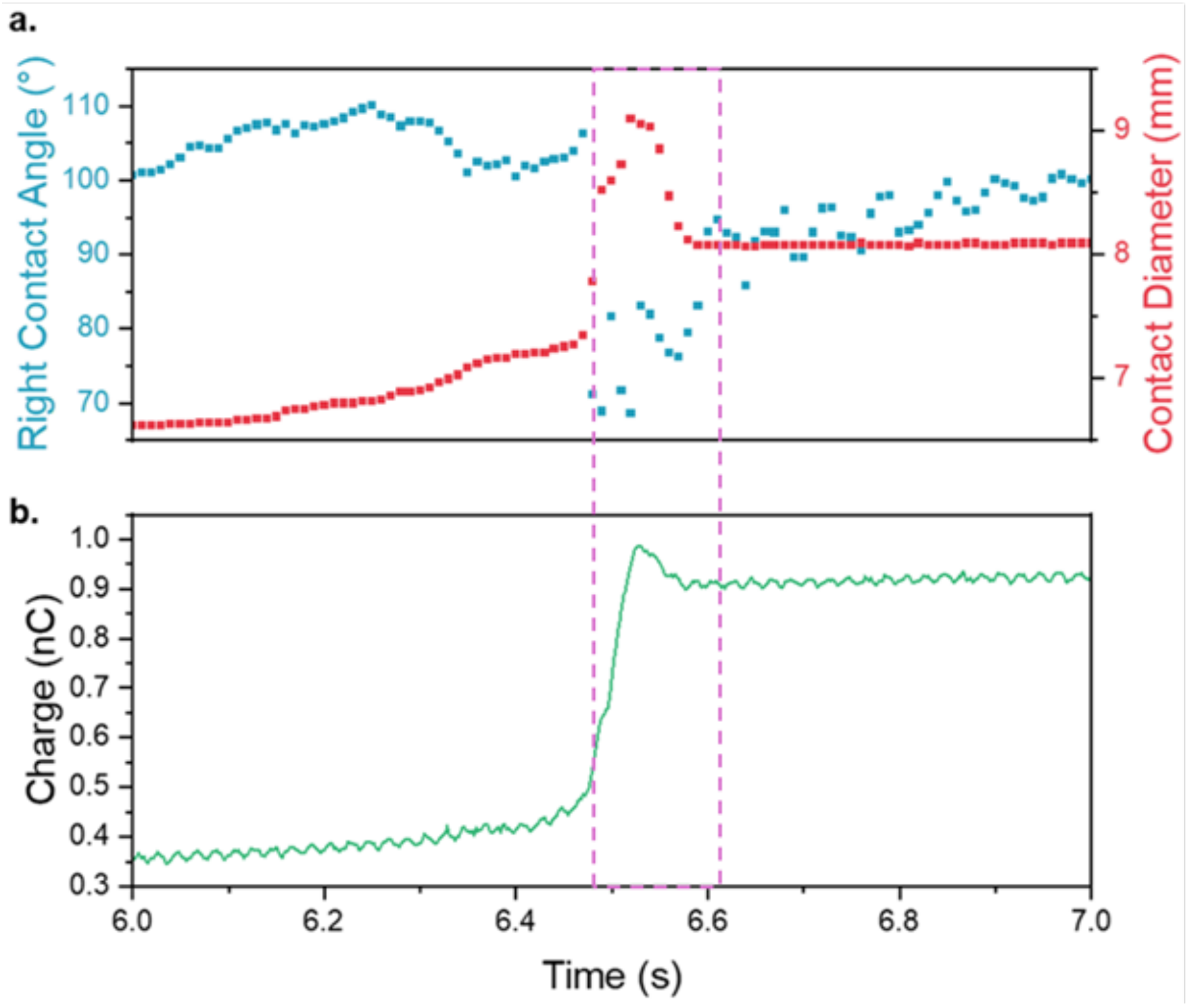

**Figure S12. a.** Contact angle measurement (blue) with corresponding variation of the liquid droplet's contact diameter (red) of the second peak during the first wetting stage; **b.** Corresponding charge measurement of the second peak during the first wetting stage.